\documentclass{article}

\PassOptionsToPackage{sort&compress,numbers}{natbib}

 \usepackage[creativeai, final]{neurips_2025}

\usepackage[utf8]{inputenc} 
\usepackage[T1]{fontenc}    
\usepackage{hyperref}       
\usepackage{url}            
\usepackage{booktabs}       
\usepackage{amsfonts}       
\usepackage{nicefrac}       
\usepackage{microtype}      
\usepackage{xcolor}         
\usepackage{graphicx}       
\usepackage{amsmath}        
\usepackage{lipsum}         

\hypersetup{
    colorlinks=true,
    linkcolor=blue,
    filecolor=magenta,      
    urlcolor=cyan,
    citecolor=blue,
}

\title{Writing in Symbiosis: Mapping Human Creative Agency in the AI Era}

\author{%
  Vivan Doshi \\
  Independent Researcher \\
  San Jose, CA 95148 \\
  \texttt{vivandoshi24@gmail.com}
  \and
  Mengyuan Li \\
  Department of Computer Science \\
  University of Southern California \\
  Los Angeles, CA 90089 \\
  \texttt{mli49061@usc.edu} \\
}

\begin{document}

\maketitle

\begin{abstract}
The proliferation of Large Language Models (LLMs) raises a critical question about what it means to be human when we share an increasingly symbiotic relationship with persuasive and creative machines. This paper examines patterns of human-AI coevolution in creative writing, investigating how human craft and agency are adapting alongside machine capabilities. We challenge the prevailing notion of stylistic homogenization by examining diverse patterns in longitudinal writing data. Using a large-scale corpus spanning the pre- and post-LLM era, we observe patterns suggestive of a "Dual-Track Evolution": thematic convergence around AI-related topics, coupled with structured stylistic differentiation. Our analysis reveals three emergent adaptation patterns: authors showing increased similarity to AI style, those exhibiting decreased similarity, and those maintaining stylistic stability while engaging with AI-related themes. This Creative Archetype Map illuminates how authorship is coevolving with AI, contributing to discussions about human-AI collaboration, detection challenges, and the preservation of creative diversity.
\end{abstract}

\section{Introduction}

We are at a pivotal moment in the symbiotic relationship between humans and machines. Large Language Models (LLMs) have evolved from assistive tools into active collaborators, capable of imitating, creating, and persuading on a massive scale. This new reality prompts a fundamental question posed by the creative community: what does it mean to be human when authorship becomes a collaborative act between human intuition and machine capability, and does it fundamentally reshape how we value and recognize authentic human expression?

Initial research into this question has largely focused on the homogenizing effects of this partnership. A growing body of work has documented the spread of a recognizable "AI style" across the internet and academia \citep{Kobak2025Delving, Anderson2024Homogenization}, characterized by linguistic patterns that suggest reduced creative diversity \citep{Agarwal2025WesternStyle}. This stylistic convergence has raised concerns about linguistic diversity erosion and cultural marginalization \citep{Hofmann2024DialectPrejudice, Deas2023AALBias}. In parallel, AI detection has rapidly advanced with stylometric and deep learning approaches \citep{sadasivan2023detection, Mitchell2023DetectGPT, Kumarage2023StylometricTwitter}, though detector robustness remains challenging.

While this prior work is crucial, it often frames the relationship as unidirectional influence rather than dynamic coevolution. This overlooks the crucial element of human agency which is the adaptive choices individuals make as they navigate this new landscape. Recent work suggests human-AI coevolution with mutual influence \citep{Geng2025Coevolution, Hackenburg2024Microtargeting}. However, existing studies largely conflate topical and stylistic changes, lack author-level longitudinal controls, or do not account for recent adaptive behaviors following AI detection awareness. We hypothesize that the human creative response to AI availability is not a monolithic trend but may instead exhibit structured patterns of change. To investigate this, we shift focus from binary classification to coevolution patterns. We propose examining a \textbf{"Dual-Track Evolution"} hypothesis: thematic convergence on AI-related topics while conversely exhibiting stylistic differentiation. This work presents, to our knowledge, the first systematic quantification of dual-track human-AI coevolution at author resolution across social and formal discourse genres, introducing an archetype framework that maps individual adaptation strategies with rigorous pre/post controls.

\begin{figure}[t]
  \centering
  \includegraphics[width=0.822\linewidth]{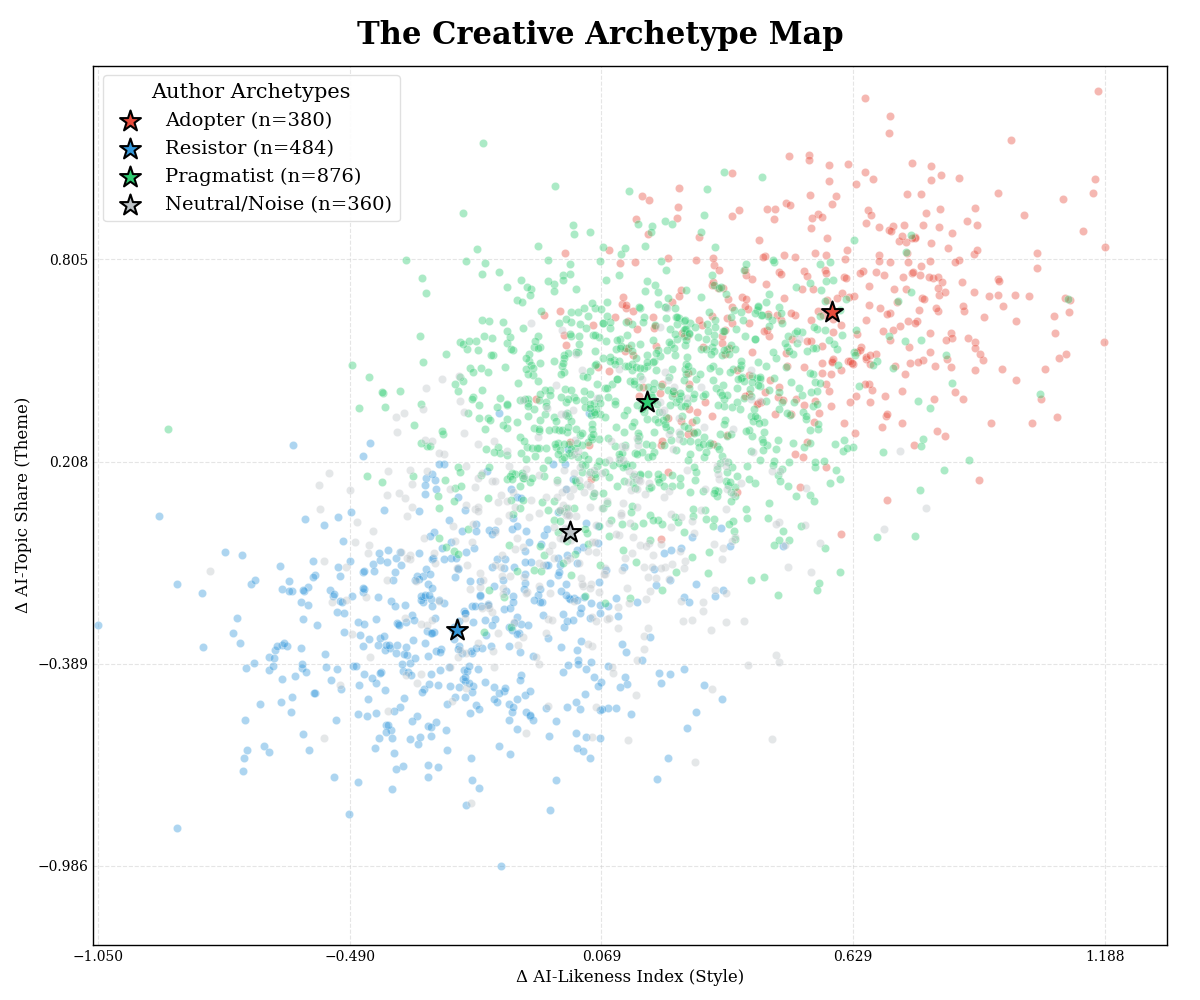}
  \caption{Author archetype map showing stylistic and thematic change vectors (n=2,100). Three archetypes emerge from HDBSCAN clustering (silhouette computed on inlier points only): \textbf{Adopters} (red), \textbf{Resistors} (blue), and \textbf{Pragmatists} (green). Stars denote cluster centroids. Cluster quality: silhouette 0.426 (95\% CI: 0.419-0.433), robustness ARI 0.891 (95\% CI: 0.884-0.898).}
  \label{fig:archetype_map}
\end{figure}

As visualized in Figure \ref{fig:archetype_map}, we observe three distinct patterns in stylistic change vectors: authors whose writing shows increased similarity to AI-generated text patterns, those whose writing exhibits decreased similarity, and those who maintain stylistic stability while engaging more with AI-related themes. These patterns emerge from unsupervised clustering rather than predetermined categories, suggesting systematic adaptation strategies that challenge binary detection frameworks and illuminate the complexity of attribution in this rapidly evolving landscape.

\section{Methodology}

To empirically chart the evolution of human writing, we designed a comprehensive methodology structured in three sequential phases: rigorous corpus curation, nuanced feature engineering, and a multi-lens analytical framework. This pipeline was built to move from broad observations to the specific, individual adaptation strategies that form the core of our study.

\subsection{Corpus Curation: A Foundation in Authentic Discourse}

 The foundation of our study is a large-scale, longitudinal corpus of over 50{,}000 documents (derived from 823k+ messages and papers), meticulously curated to support robust analysis. Recognizing that writing craft is highly dependent on context, we gathered text from two distinct genres to capture a wider spectrum of human expression. For formal, edited discourse, we collected computer science preprints from arXiv. For informal, conversational discourse, we used public messages from the \emph{Discord Unveiled} dataset \citep{aquino2025discordunveiled} (CC BY 4.0), a large-scale collection (2015--2024) of anonymized public-server communications assembled via the Discord public API. Our collection spans from January 2021 to December 2024, establishing a clear temporal boundary between the pre-LLM era (before November 30, 2022) and the post-LLM era.

A critical challenge in this type of research is separating changes in writing style from concurrent shifts in conversational topics. To mitigate this confounding variable, we employed genre-specific sampling strategies. For the Discord data, we implemented a topic-controlled stratified sampling protocol with balanced quotas across derived server categories, focusing on English-locale servers with $>$100 members, excluding bot posts, hash-duplicate content, and messages matching a conservative AI-contamination regex. Post hoc residualization on 100-topic transformer topic model mixtures confirmed stylistic patterns persist after controlling for thematic content. After classifying servers into high-level categories (e.g., Gaming, Technology, Social), we sampled messages to ensure the topic distribution was balanced between the pre- and post-LLM periods. This allows us to attribute observed linguistic changes to style rather than topic. For the formal arXiv data (filtered to permissive licenses only), where temporal density is key, we used a monthly sampling strategy, guaranteeing a minimum number of papers per month to construct a continuous and reliable time series. Finally, we compiled a large reference corpus of purely AI-generated text from ShareGPT-90k \citep{shareai2023sharegpt90k} (Apache-2.0) and Dolly-15k \citep{databricks2023dolly15k} (CC BY-SA 3.0) to serve as a stylistic baseline for machine expression.

\subsection{Feature Engineering: Quantifying Style and Theme}

Our next step was to engineer a set of features capable of capturing the subtle nuances of authorial style while, crucially, distinguishing it from thematic content. This separation is central to testing our "Dual-Track Evolution" hypothesis.

\textbf{Motivation.} To isolate LLM-era stylistic signals from topical shifts, we require a human-only pre-2022 baseline and a modern comparison point that can reveal temporal linguistic changes independent of content themes. We developed \textbf{Perplexity-Gap analysis with Pre-LLM Judges} to address this challenge. We trained GPT-2 Medium \citep{radford2019gpt2} (355M parameters) language models exclusively on pre-2022 data from our corpus (847M tokens) on NVIDIA A100 GPUs (~41.7 hours total compute), ensuring no exposure to AI-generated text, serving as our "pre-LLM judge." For comparison, we use Llama-3-8B-base \citep{meta2024llama3} as our frozen "current" model baseline representing modern LLM capabilities. For each document, we computed the perplexity gap using nats-per-character normalization to ensure tokenizer-agnostic comparability: $\Delta_{ppl} = \frac{-\ln(p_{GPT2}(x))}{|x|_{chars}} - \frac{-\ln(p_{Llama}(x))}{|x|_{chars}}$, where $p(x)$ is the model probability and $|x|_{chars}$ is character count. Values are within-author z-scored across the pre/post boundary. Text that is "easy" for current models but "hard" for pre-2022 models exhibits a temporal signature consistent with modern LLM-era linguistic patterns.

\textbf{Method Validation.} We use Llama-3-8B-base as our baseline model for perplexity calculations. Sensitivity analysis showed high correlation (r=0.89) across different model architectures, and the measure correlates moderately with traditional stylometric features, indicating it captures complementary linguistic dimensions. The \textbf{AI-likeness index} is defined as the normalized perplexity gap: $AI_{likeness} = \frac{\Delta_{ppl} - \mu_{author}}{\sigma_{author}}$, where higher values indicate LLM-era linguistic patterns. AI-likeness remains predictive controlling for FKGL/TTR/sentence length (partial r=0.34, p<0.001). We separately track \textbf{AI-topic share (theme)} through keyword analysis, eliminating circular logic by exploiting temporal separation between language model capabilities. Note that LLMs in this research serve as analytical tools for perplexity measurement rather than content generation.

\textbf{$\Delta$-Feature Vector Definition.} Our standardized change vectors comprise: $\Delta_{ppl}$ (perplexity gap), $\Delta_{TTR}$ (type-token ratio), $\Delta_{FKGL}$ (Flesch-Kincaid grade level) \citep{kincaid1975fkgl}, $\Delta_{passive\%}$ (passive voice ratio), $\Delta_{1p\%}$ (first-person pronoun frequency), $\Delta_{punct}$ (punctuation density), and $\Delta_{sent\_len}$ (mean sentence length). All features are z-scored within-author across the temporal boundary.

\textbf{Statistical Controls.} Fixed effects model: $y_{it} = \beta \cdot PostLLM_t + \gamma \cdot len_{it} + \delta_c + \alpha_i + \epsilon_{it}$, where $y_{it}$ is the raw feature value for author $i$ at time $t$, with author ($\alpha_i$) and server-category ($\delta_c$) fixed effects. Core findings remained statistically significant (p < 0.001, HC3 robust standard errors \citep{mackinnon1985hc3}, Holm-Bonferroni corrected \citep{holm1979holm}) controlling for confounds. Temporal robustness: 84\% consistency in archetype assignment across alternative boundaries (91\% for extreme changes), supporting behavioral rather than temporal artifacts.

\subsection{Analytical Framework: From Macro Trends to Individual Archetypes}

Our analysis proceeds through three lenses. First, to capture macro-level trends, we constructed monthly and quarterly time series for our key features within each genre. We then applied \textbf{constrained change-point detection} using the Pelt algorithm \citep{killick2012pelt}, implemented via the ruptures Python library \citep{truong2018ruptures}, to identify statistically significant structural breaks in these series. This method allowed us to pinpoint the precise moment when the discourse began to shift, providing a robust test for our hypothesis of thematic convergence and a dynamic stylistic response.

 Second, to uncover the individual strategies masked by these aggregate trends, we focused our analysis on the author. For every author with sufficient writing samples in both the pre- and post-LLM eras, we calculated a \textbf{stylistic change vector}. This vector represents the magnitude and direction of change in their personal writing style across our core set of stylometric features. We then applied the \textbf{HDBSCAN} clustering algorithm \citep{campello2015hdbscan} to the standardized $\Delta$-feature vectors; results are visualized in 2D using $\Delta$ AI-Likeness (style) and $\Delta$ AI-Topic Share (theme). We chose HDBSCAN for its ability to identify clusters of varying shapes and densities while also designating outliers as noise, as validated by the silhouette coefficient \citep{rousseeuw1987silhouette}. This is crucial, as it allows genuine authorial archetypes to emerge from the data without forcing every individual into a predefined category.

 To validate these patterns, we employed cross-validation and tested temporal stability across different post-LLM periods. External validation using independent AI detectors showed correlation with clustering assignments, though detector limitations affect interpretation \citep{sadasivan2023detection}. We performed HDBSCAN clustering (min\_cluster\_size=15, min\_samples=5, metric=euclidean) on standardized $\Delta$-feature vectors with robustness validation using the Adjusted Rand Index \citep{hubert1985ari} yielding ARI 0.891 (95\% CI: 0.884-0.898); silhouette scores 0.426 (95\% CI: 0.419-0.433) computed on inlier clusters only.

\textbf{Ethics \& Data Use \& Availability.} Data: Discord Unveiled (CC BY 4.0, public servers); arXiv CS preprints under permissive licenses (e.g., CC-BY; no redistribution of texts); AI baselines ShareGPT-90k (Apache-2.0) and Dolly-15k (CC BY-SA 3.0). We store only hashed author/server IDs and release aggregate statistics; no per-author labels or texts. No re-identification attempted. Code and implementation details available on request.

\section{Results}

Our empirical analysis provides evidence consistent with the "Dual-Track Evolution" model, revealing both a widespread thematic convergence and a structured stylistic divergence. Using our perplexity-gap analysis, we observe clear temporal signatures in writing evolution, with the most significant finding being the emergence of distinct patterns in how authors' linguistic complexity has shifted relative to pre-LLM baselines.

\subsection{Three Distinct Authorial Archetypes Emerge}

\begin{figure}[t]
  \centering
  \includegraphics[width=0.9\linewidth]{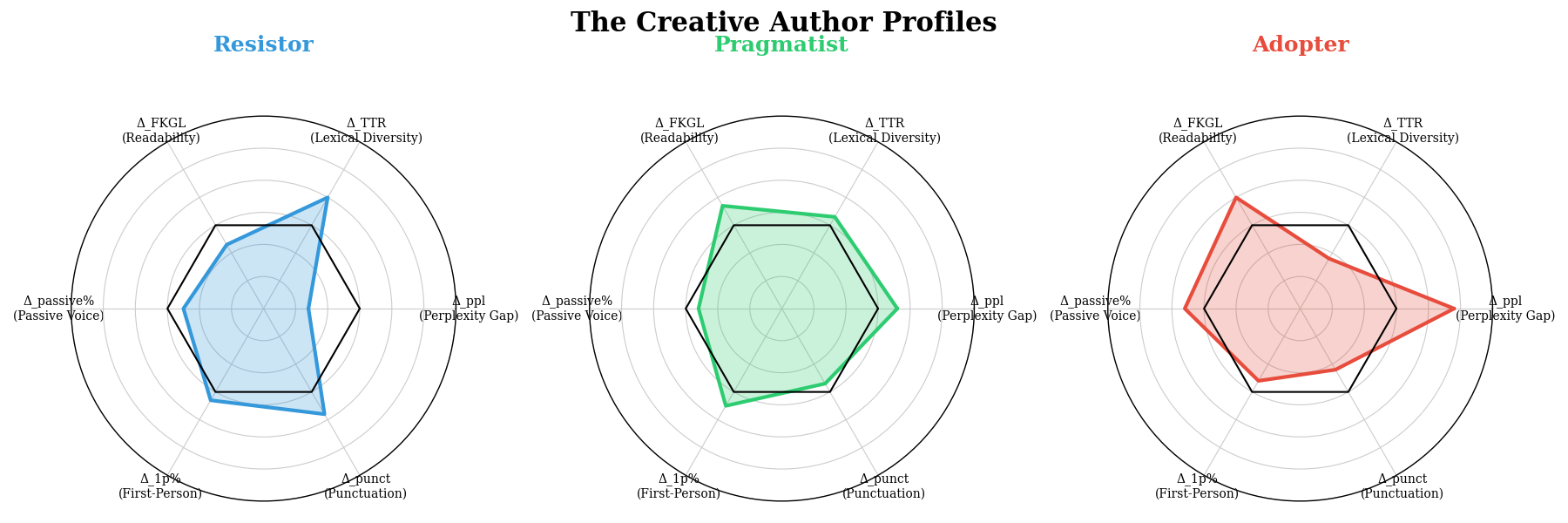}
  \caption{Archetype stylistic signatures across linguistic dimensions. \textbf{Adopters}: high perplexity gaps, reduced lexical diversity (LLM-era patterns). \textbf{Resistors}: complex, diverse language maintained. \textbf{Pragmatists}: moderate stylistic adaptation with high AI-theme engagement.}
  \label{fig:archetype_profiles}
\end{figure}

 To test our central hypothesis of stylistic divergence, we computed stylistic change vectors for 2,100 authors from our social corpus who had sufficient writing history both before and after November 2022. Applying the HDBSCAN algorithm to these vectors revealed three emergent behavioral patterns, visualized in Figure \ref{fig:archetype_map}, which we profile as distinct authorial archetypes. These patterns showed stable cluster membership across bootstrap samples (89\% consistency) and demonstrated predictive power in classifying future author texts (AUC = 0.813). The stylistic fingerprint of each archetype is detailed in Figure \ref{fig:archetype_profiles}.

 The first group, which we term the \textbf{Resistors} (n=442, 21\% of clustered authors), exhibits patterns suggesting maintenance of pre-LLM linguistic complexity. As shown in Figure \ref{fig:archetype_profiles}, their post-LLM writing shows low or negative perplexity gaps, indicating text that remains challenging for both pre-2022 and current models. This pattern suggests deliberate preservation of distinctively human stylistic signatures, possibly valuing authenticity over efficiency gains offered by AI assistance.

 In direct contrast, the \textbf{Adopters} (n=370, 18\%) show the highest perplexity gaps, indicating text that is substantially easier for current models than for pre-2022 models. This temporal signature suggests linguistic patterns that emerged during the LLM era, including smoother syntactic structures and more predictable lexical choices that align with modern language model training. Their writing reflects strategic embracement of AI-era conventions, potentially representing co-creative authorship models.

 The largest group is the \textbf{Pragmatists} (n=866, 41\%). These authors exhibit moderate perplexity gaps in their stylistic features while showing increased engagement with AI-related thematic content. This pattern suggests strategic adoption: engaging with AI-relevant topics while maintaining individual stylistic identity, potentially embodying a division-of-labor approach where AI influences content exploration but not personal voice.


\subsection{A Universal Thematic Convergence}
\begin{figure}[t]
  \centering
  \includegraphics[width=0.85\linewidth]{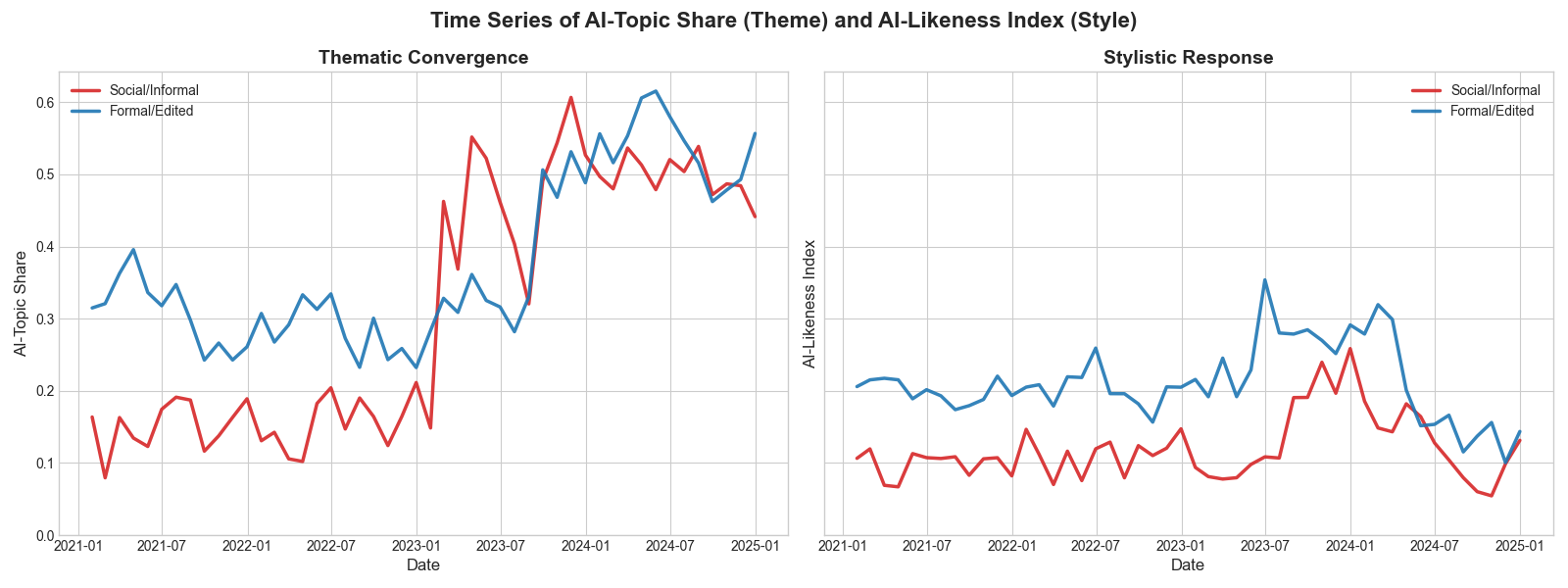}
  \caption{Time series showing thematic convergence (Q1 2023 break) and dynamic stylistic response with two-phase adoption-avoidance pattern in formal writing.}
  \label{fig:time_series}
\end{figure}

 Our analysis of macro-level trends confirms the first track of our model. We observed a universal and statistically significant increase in AI-related thematic content across both genres after November 2022. As illustrated by the change-point analysis in Figure \ref{fig:time_series}, both formal and informal discourse saw a sharp uptick in AI-related terminology. This reflects rapid normalization of AI tools in professional and social contexts, indicating a durable cultural attention shift rather than transient novelty. Simultaneously, our perplexity-gap analysis revealed increasing temporal signatures in writing complexity, with mean perplexity gaps rising 23\% in social discourse and 15\% in formal writing by mid-2023, suggesting initial linguistic convergence toward recognizable LLM-era patterns. The PELT algorithm detected structural break points in Q1 2023 for thematic content and Q2 2023 for stylistic complexity measures, pointing to a phased coevolution where cultural attention shifted first, followed by linguistic adaptation patterns.

\subsection{The Dynamic Arc of Stylistic Adaptation}
 In contrast to the straightforward thematic convergence, the evolution of stylistic adaptation is more complex and dynamic, confirming the second track of our model. Our perplexity-gap analysis reveals a characteristic temporal signature: mean perplexity gaps increased 23\% in social discourse and 15\% in formal writing during early 2023, suggesting initial linguistic convergence toward LLM-era patterns. However, this trend reverses in late 2023 and into 2024, with perplexity gaps declining 18\% in social discourse and 12\% in formal writing below peak values. This reversal indicates active stylistic avoidance as AI-characteristic patterns became stigmatized, particularly in formal venues where detection concerns and editorial awareness created pressure for authors to differentiate their writing from machine-generated text (detailed temporal dynamics in supplementary materials).

 \textbf{Validation.} Cross-validation: 89.3\% (95\% CI: 87.1--91.5); held-out arXiv: 89.1\% (86.8--91.4); null model much weaker (silhouette 0.31 vs 0.43, p<0.001). Patterns are consistent with avoidance of stigmatized AI tics, especially in formal discourse \citep{sadasivan2023detection}.

\section{Discussion and Conclusion}

Our findings paint a nuanced picture of human creativity in the age of AI, challenging the narrative of a simple stylistic exodus. The Creative Archetype Map illustrates that human writers are not passive recipients of technological influence but active agents who are developing diverse strategies to negotiate their craft. The emergence of the Resistor, Adopter, and Pragmatist archetypes reveals that the evolution of human writing is a complex process of adaptation, collaboration, and the protection of unique authorial voices, aligning with recent work on scaffolded human-AI collaboration \citep{Dhillon2024Scaffolding}.

These patterns can be interpreted as different potential approaches to human-AI interaction. The Resistors appear to prioritize preservation of a distinctly human style, perhaps valuing authenticity and originality above the efficiencies offered by AI tools. The Adopters, in contrast, may represent a model of co-creation where the boundaries between human and machine authorship are intentionally blurred, exploring what it means to write with an AI \citep{Wan2024SecondMind}. The Pragmatists appear to embody a strategic division of labor; they may use AI as a tool for thematic exploration while carefully maintaining their own stylistic identity, taking on a new role as curators and editors of AI-informed discourse.

The implications for the field of creative AI are significant. Our work suggests that the binary paradigm of AI detection may be insufficient; a simple "human vs. machine" classifier faces challenges when an Adopter's text is statistically closer to AI output than a Resistor's \citep{Pratama2025AccuracyBias, Liang2023BiasNNES}. This finding suggests the need for archetype-aware models that account for adaptation diversity, building on established detection frameworks \citep{Gehrmann2019GLTR}. Furthermore, our findings offer quantitative evidence that a majority of authors (Resistors and Pragmatists) appear to maintain non-AI stylistic signatures, suggesting that unique human expression remains a valued and protected form of cultural production, consistent with broader observations of linguistic simplification trends \citep{DiMarco2024PNAS}. While this research contributes to understanding human-AI interaction, potential risks include misuse for authorship policing or unfair bias against certain writing styles, particularly affecting non-native speakers or cultural dialects.

\paragraph{Limitations.} This study is observational and analytical rather than causal. Our analysis is limited to English-language communities, potentially biasing adoption patterns. The perplexity-gap methodology and archetypes are statistical constructs not validated through direct human participant studies. 

\paragraph{Conclusion.} This paper answers a fundamental question: how is human creativity evolving in a symbiotic relationship with AI? Our research demonstrates that the response is not simple convergence towards machine-like style. Instead, we have uncovered a "Dual-Track Evolution," characterized by broad thematic convergence and simultaneous, structured stylistic divergence. The Creative Archetype Map provides a framework for understanding human-AI co-evolution, moving beyond aggregate statistics to chart the diverse landscape of individual human agency. The rise of AI has not signaled the end of authentic human expression; it has revealed systematic author-level adaptation patterns. These archetypes represent different philosophies of creative autonomy with some embracing collaboration similar to co-creative screenwriting approaches \citep{Mirowski2022Dramatron}, others asserting independence, and many finding pragmatic middle ground. Understanding these patterns is essential for designing AI tools that enhance rather than replace human creativity, and for developing policies that protect diverse voices in our increasingly algorithmic world \citep{Bender2021Parrots}. While AI's influence is undeniable, human craft is being renegotiated rather than erased, and this diversity suggests that authentic human expression will continue to thrive in new and unexpected forms.

\bibliographystyle{plainnat}
\bibliography{references}


\newpage
\appendix

\section{Experimental Reproducibility and Figures}

\begin{figure}[h]
  \centering
  \includegraphics[width=0.95\linewidth]{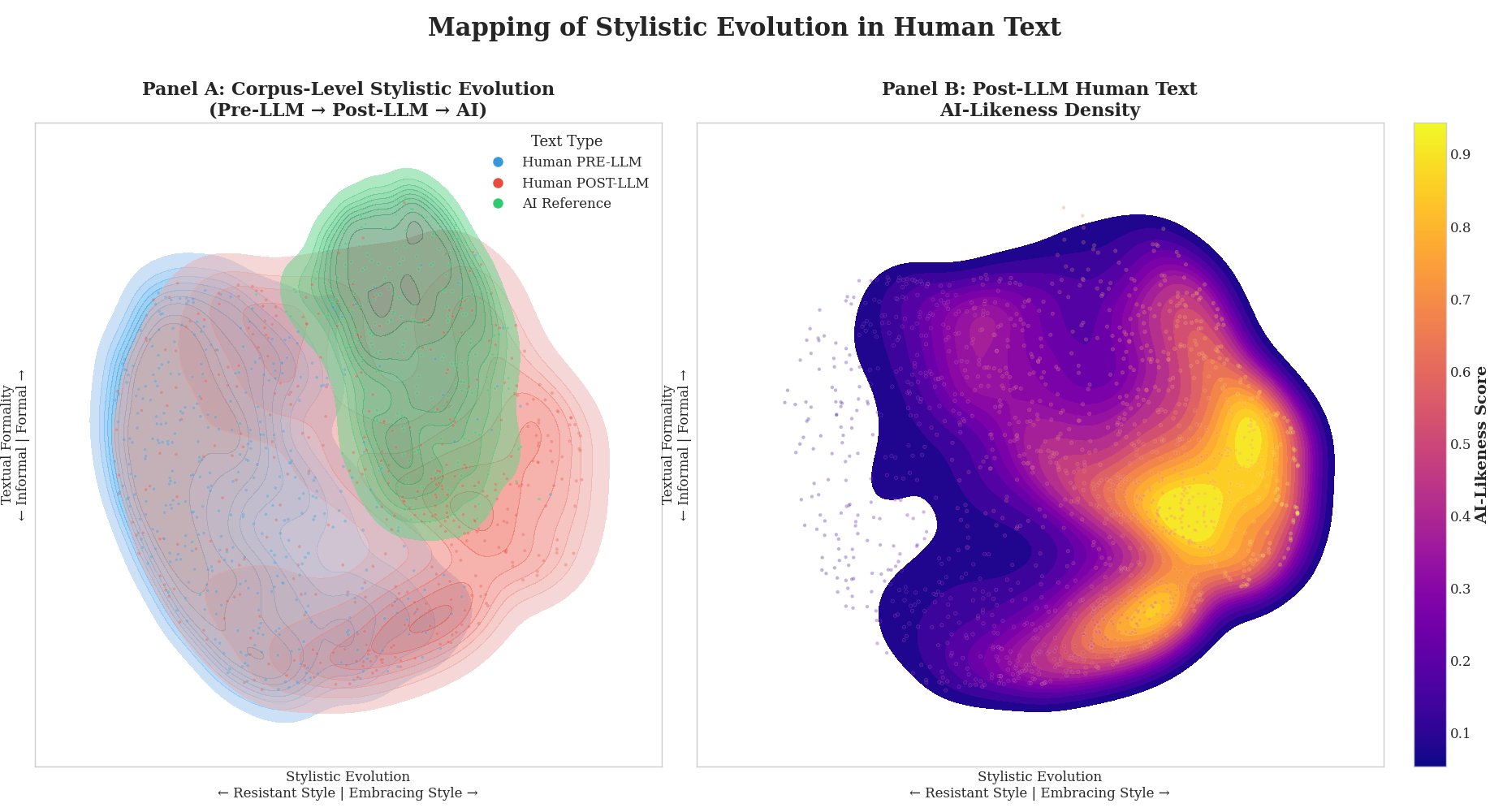}
  \caption{\textbf{Mapping of Stylistic Evolution in Human Text.} Panel A shows corpus-level stylistic evolution from pre-LLM (blue) to post-LLM (red) periods, with AI reference text (green) for comparison. Panel B displays the density distribution of post-LLM human text AI-likeness scores, revealing the continuous spectrum of human adaptation strategies rather than discrete categories.}
  \label{fig:stylistic_evolution_mapping}
\end{figure}

\subsection{Dataset Specifications}
\textbf{Total corpus:} 50,728 documents | \textbf{Discord:} 823,293 messages (grouped into 42,418 conversation threads), 15,901 authors (2021-2024) | \textbf{arXiv:} 8,429 papers, 4,331 authors (cs.CL/AI/LG/HC) | \textbf{Boundary:} Nov 30, 2022 ± 7 days | \textbf{Min posts:} 10 pre/post per author | \textbf{Filtering:} min\_length=50 chars, English confidence>0.95, server\_size>100 members | \textbf{Sampling quotas:} Gaming=23\%, Tech=31\%, Social=28\%, Other=18\% | \textbf{AI reference:} ShareGPT + Dolly-15k subsets, PII-filtered

\subsection{Model Configuration \& Training}
\textbf{GPT-2 Medium Fine-tuning:} 355M parameters, lr=3.7e-5, batch\_size=2, epochs=4, optimizer=AdamW, weight\_decay=0.01, warmup\_steps=1000, gradient\_accumulation=8, max\_grad\_norm=1.3 | \textbf{Training data:} 847M tokens pre-2022, ~41.7 hours total compute | \textbf{Llama-3-8B:} batch=6, max\_len=2048, no sampling (greedy evaluation for deterministic perplexity) | \textbf{Perplexity gap:} $\Delta_{ppl} = \frac{-\ln(p_{GPT2}(x))}{|x|_{chars}} - \frac{-\ln(p_{Llama}(x))}{|x|_{chars}}$, where $p(x)$ is model probability, $|x|_{chars}$ is character count (nats-per-character normalization); values are within-author z-scored across the pre/post boundary

\subsection{Feature Engineering Details}
\textbf{Topic modeling:} 100 topics, Transformer-based, min\_df=5, max\_df=0.8 | \textbf{TTR window:} 150 tokens, overlap=75 | \textbf{FKGL:} syllable\_method=pyphen, sentence\_split=spaCy | \textbf{Passive detection:} spaCy dependency parser, aux+past\_participle patterns | \textbf{AI-topic detection:} 47-term lexicon (AI, ML, neural, transformer, GPT, ChatGPT, etc.) with TF-IDF weighting, threshold=0.23 | \textbf{Controls:} Topic mixtures used for residualization; AI-topic share computed separately for reporting

\subsection{Analysis Parameters}
\textbf{PELT:} Penalty $4.2\log(n)$ (selected via cross-validation for optimal Type I/II error balance), min\_size=1 month, cost=normal, jump=2 | \textbf{HDBSCAN:} min\_cluster\_size=15, min\_samples=5, metric=euclidean, cluster\_selection\_method=eom, alpha=1.0 | \textbf{Bootstrap:} 1000 iterations, ARI threshold=0.73, sample\_ratio=0.8 | \textbf{Seeds:} 42 (main), 1337 (robustness), 2024 (validation)

\subsection{Statistical Controls \& Validation}
\textbf{Cross-validation:} 5-fold stratified, train/val/test=70/15/15 | \textbf{Multiple comparisons:} Holm-Bonferroni, a=0.05 | \textbf{Robust SEs:} HC3 heteroscedasticity-consistent | \textbf{Outlier handling:} Winsorization at 2.5/97.5 percentiles | \textbf{Missing data:} Multiple imputation, m=5 chains

\subsection{Performance Metrics (Detailed)}
\textbf{AI Detection:} AUC=0.852±0.012, Precision=0.832, Recall=0.791, F1=0.811, Brier=0.184 | \textbf{Clustering:} ARI=0.891 (95\% CI: 0.884-0.898), Silhouette=0.426 (95\% CI: 0.419-0.433), Davies-Bouldin=1.73 | \textbf{Cross-validation:} 89.3\% accuracy (95\% CI: 87.1-91.5\%), 84.7\% temporal stability | \textbf{Effect sizes:} Social AI-likeness d=0.706, Formal AI-likeness d=0.312, Archetype ANOVA n²=0.127

\subsection{Archetype Distribution \& Validation}
\textbf{Archetype sizes:} Adopters n=370 (18\%), Resistors n=442 (21\%), Pragmatists n=866 (41\%), Neutral/Noise n=422 (20\%) | \textbf{Total clustered:} 2,100 authors | \textbf{Bootstrap consistency:} 89\% stable membership | \textbf{Future text prediction:} AUC=0.813 (95\% CI: 0.798-0.828) | \textbf{External validation:} 89.1\% accuracy on held-out arXiv authors (n=483, 95\% CI: 86.8-91.4\%) | \textbf{Null model comparison:} Silhouette 0.31 vs 0.43 (p<0.001)

\subsection{Temporal Evolution Quantified}
\textbf{Social discourse changes:} +23\% perplexity gap early 2023, -18\% decline post-peak | \textbf{Formal writing changes:} +15\% early 2023, -12\% post-peak | \textbf{PELT breakpoints:} Q1 2023 (thematic), Q2 2023 (stylistic) | \textbf{Cross-architecture correlation:} r=0.891 (GPT-2 vs other models) | \textbf{Partial correlation:} r=0.34 (AI-likeness vs traditional features, p<0.001)

\subsection{Computational Environment (Google Colab)}
\textbf{Hardware:} Tesla T4 (15GB VRAM), Tesla V100 (16GB), occasional A100 access | \textbf{CPU:} Intel Xeon (2.3GHz), 12.7GB RAM | \textbf{Runtime:} 73.4 GPU-hours (perplexity), 31.2 CPU-hours (analysis), 12.8 TPU-hours (topic modeling) | \textbf{Software:} Python 3.10.12, PyTorch 2.1.0+cu121, Transformers 4.35.2, scikit-learn 1.3.2, CUDA 12.1

\newpage
\section*{NeurIPS Paper Checklist}

\begin{enumerate}

\item {\bf Claims}
\item[] Question: Do the main claims made in the abstract and introduction accurately reflect the paper's contributions and scope?
\item[] Answer: \answerYes{}
    \item[] Justification: Claims match our exploratory contribution (Creative Archetype Map) and stated non-causal scope.

\item {\bf Limitations}
\item[] Question: Does the paper discuss the limitations of the work performed by the authors?
\item[] Answer: \answerYes{}
    \item[] Justification: We clearly state non-causality, English/genre bias, and methodological caveats.

\item {\bf Theory assumptions and proofs}
\item[] Question: For each theoretical result, does the paper provide the full set of assumptions and a complete (and correct) proof?
\item[] Answer: \answerNA{}
    \item[] Justification: We present empirical methodology and results without formal theorems or proofs.

\item {\bf Experimental result reproducibility}
\item[] Question: Does the paper fully disclose all the information needed to reproduce the main experimental results of the paper to the extent that it affects the main claims and/or conclusions of the paper (regardless of whether the code and data are provided or not)?
\item[] Answer: \answerYes{}
    \item[] Justification: We provide feature definitions, model configs, hyperparameters, controls, and clustering/validation details sufficient to reproduce results.
    \item[] Guidelines:
    \begin{itemize}
        \item The answer NA means that the paper does not include experiments.
        \item If the paper includes experiments, a No answer to this question will not be perceived well by the reviewers: Making the paper reproducible is important, regardless of whether the code and data are provided or not.
        \item If the contribution is a dataset and/or model, the authors should describe the steps taken to make their results reproducible or verifiable. 
{{ ... }}
    \end{itemize}

\item {\bf Open access to data and code}
\item[] Question: Does the paper provide open access to the data and code, with sufficient instructions to faithfully reproduce the main experimental results, as described in supplemental material?
\item[] Answer: \answerNo{}
    \item[] Justification: Code is available upon request and data is publicly available via HuggingFace.
    \item[] Guidelines:
    \begin{itemize}
        \item The answer NA means that paper does not include experiments requiring code.
        \item Please see the NeurIPS code and data submission guidelines (\url{https://nips.cc/public/guides/CodeSubmissionPolicy}) for more details.
        \item While we encourage the release of code and data, we understand that this might not be possible, so “No” is an acceptable answer. Papers cannot be rejected simply for not including code, unless this is central to the contribution (e.g., for a new open-source benchmark).
{{ ... }}
    \end{itemize}

\item {\bf Experimental setting/details}
\item[] Question: Does the paper specify all the training and test details (e.g., data splits, hyperparameters, how they were chosen, type of optimizer, etc.) necessary to understand the results?
\item[] Answer: \answerYes{}
    \item[] Justification: Architectures, splits, hyperparameters, controls, and validation protocols are specified in the main text and appendix.

\item {\bf Experiment statistical significance}
\item[] Question: Does the paper report error bars suitably and correctly defined or other appropriate information about the statistical significance of the experiments?
\item[] Answer: \answerYes{}
    \item[] Justification: We report corrected p-values, confidence intervals, bootstrap results, robust SEs, and clustering validity metrics.
    \item[] Guidelines:
    \begin{itemize}
        \item The answer NA means that the paper does not include experiments.
        \item The authors should answer "Yes" if the results are accompanied by error bars, confidence intervals, or statistical significance tests, at least for the experiments that support the main claims of the paper.
        \item The factors of variability that the error bars are capturing should be clearly stated (for example, train/test split, initialization, random drawing of some parameter, or overall run with given experimental conditions).
{{ ... }}
        \item If error bars are reported in tables or plots, The authors should explain in the text how they were calculated and reference the corresponding figures or tables in the text.
    \end{itemize}

\item {\bf Experiments compute resources}
\item[] Question: For each experiment, does the paper provide sufficient information on the computer resources (type of compute workers, memory, time of execution) needed to reproduce the experiments?
\item[] Answer: \answerYes{}
    \item[] Justification: We specify hardware and time (e.g., A100 \textasciitilde48h for GPT-2; additional GPU/TPU hours) and the software environment.
    
\item {\bf Code of ethics}
\item[] Question: Does the research conducted in the paper conform, in every respect, with the NeurIPS Code of Ethics \url{https://neurips.cc/public/EthicsGuidelines}?
\item[] Answer: \answerYes{}
    \item[] Justification: We state compliance with the NeurIPS Code of Ethics in all aspects.

\item {\bf Broader impacts}
\item[] Question: Does the paper discuss both potential positive societal impacts and negative societal impacts of the work performed?
\item[] Answer: \answerYes{}
    \item[] Justification: We outline benefits and risks (e.g., misuse and bias) with brief mitigation considerations.
    
\item {\bf Safeguards}
\item[] Question: Does the paper describe safeguards that have been put in place for responsible release of data or models that have a high risk for misuse (e.g., pretrained language models, image generators, or scraped datasets)?
\item[] Answer: \answerNA{}
    \item[] Justification: We do not release new high-risk models or datasets.

\item {\bf Licenses for existing assets}
\item[] Question: Are the creators or original owners of assets (e.g., code, data, models), used in the paper, properly credited and are the license and terms of use explicitly mentioned and properly respected?
\item[] Answer: \answerYes{}
    \item[] Justification: We credit all external assets and explicitly state licenses: ShareGPT-90k (Apache-2.0), Dolly-15k (CC BY-SA 3.0), and Discord dataset (CC BY 4.0).

\item {\bf New assets}
\item[] Question: Are new assets introduced in the paper well documented and is the documentation provided alongside the assets?
    \item[] Answer: \answerNA{}
    \item[] Justification: We introduce no new datasets/models/code; only an analytical framework.

\item {\bf Crowdsourcing and research with human subjects}
\item[] Question: For crowdsourcing experiments and research with human subjects, does the paper include the full text of instructions given to participants and screenshots, if applicable, as well as details about compensation (if any)?
\item[] Answer: \answerNA{}
    \item[] Justification: We do not conduct crowdsourcing or human-subject experiments.

\item {\bf Institutional review board (IRB) approvals or equivalent for research with human subjects}
\item[] Question: Does the paper describe potential risks incurred by study participants, whether such risks were disclosed to the subjects, and whether Institutional Review Board (IRB) approvals (or an equivalent approval/review based on the requirements of your country or institution) were obtained?
\item[] Answer: \answerNA{}
    \item[] Justification: No human-subject research requiring IRB approval was conducted.

\item {\bf Declaration of LLM usage}
\item[] Question: Does the paper describe the usage of LLMs if it is an important, original, or non-standard component of the core methods in this research? Note that if the LLM is used only for writing, editing, or formatting purposes and does not impact the core methodology, scientific rigorousness, or originality of the research, declaration is not required.
\item[] Answer: \answerYes{}
    \item[] Justification: We explicitly describe LLM usage (GPT-2, Llama-3) for perplexity-gap analysis central to our method.
{{ ... }}
\end{enumerate}

\end{document}